\documentclass[twoside]{article}
\usepackage{qic,epsfig}

\usepackage{bm}
\usepackage{bbm}

\newcommand{\ket}[1]{|#1\rangle}

\begin{document}
\setlength{\textheight}{8.0truein}    

\runninghead{
Entanglement probability distribution of bi-partite ...} 
            {O.C.O. Dahlsten and M.B. Plenio}

\normalsize\textlineskip
\thispagestyle{empty}
\setcounter{page}{527}


\vspace*{0.88truein}

\alphfootnote

\fpage{527}

\centerline{\bf
ENTANGLEMENT PROBABILITY DISTRIBUTION }
\vspace*{0.035truein}
\centerline{\bf OF BI-PARTITE RANDOMISED STABILIZER STATES}
\vspace*{10pt}
\centerline{\footnotesize
OSCAR C.O. DAHLSTEN\footnote{oscar.dahlsten@imperial.ac.uk}
~~and~~MARTIN B. PLENIO}
\vspace*{0.015truein}
\centerline{\footnotesize\it Institute for Mathematical Sciences, Imperial College
London} 
\baselineskip=10pt
\centerline{\footnotesize\it London SW7 2PE, UK}
\centerline{\footnotesize\it and}
\centerline{\footnotesize\it QOLS, Blackett
Laboratory, Imperial College London} 
\baselineskip=10pt
\centerline{\footnotesize\it London SW7 2BW, UK}

\vspace*{0.225truein}

\vspace*{0.21truein}

\abstracts{We study the entanglement properties of random pure stabilizer
states in spin-$\frac{1}{2}$ particles. We obtain a compact and {\em exact}
expression for the probability distribution of the entanglement values
across any bipartite cut. This allows for exact derivations
of the average entanglement and the degree of concentration of
measure around this average. We also give simple bounds on these
quantities. We find that for large systems the average entanglement
is near maximal and the measure is concentrated around it.
}{}{}
\vspace*{10pt}

\vspace*{3pt}

\vspace*{1pt}\textlineskip    
\section{Introduction}
\noindent
Entanglement is a fundamental resource for quantum information
processing. The classification and quantification of this
resource in two- and many-partite systems is therefore of
significant concern, as can be seen in
the review papers \cite{Plenio V 98, Eisert P 03,Plenio V 05,Eisert G 05,Horodecki1,Wootters,Horodecki2}.
Over the past $10$ years or so the properties of bi-partite entanglement have been
explored in some detail and many of its basic features are now
reasonably well understood. However, the entanglement properties of
multipartite systems are far more complex and
our current understanding of this setting is limited.

There are various approaches that one might take to achieve
progress. Firstly, one may impose additional constraints on the
set of states and/or the set of operations that one is interested
in, ideally without significantly reducing the variety of possible qualitative
entanglement structures. In this context an interesting
class of states that arises is that of stabilizer states.
Despite having various restrictions, these stabilizer states
possess a rich entanglement
structure exhibiting multi-partite entanglement\cite{gottesman1,gottesman2,Audenaert P 05,gottesman3,fattal,Hein EB 04}.

With increasing numbers of particles the finest classification of
entanglement results in a number of types of entanglement that
grows exponentially with the number of particles. However, many of
these types may be atypical in that their probability, with respect to some natural measure on the set of states,
vanishes in the limit of large numbers of particles. This
suggests a different approach that studies only the entanglement properties of
typical states. Even before the emergence of quantum
information such questions had already been of interest. An
example is the study of the expected entropy
\cite{lubkin,page,foong} of a subsystem when averaged over the
invariant measure of pure states. In quantum information this entropy has received considerable recent
attention as it is precisely the entanglement between
this subset and the remainder of the system. The mean entanglement has been studied, as well as the probability
for deviations from this mean. These can be shown to be
exponentially decreasing with the difference from the mean, a
property known as 'concentration of measure' \cite{hayden}.

In the present paper we are interested in a combination of
these two approaches, i.e. we study the typical and average
entanglement of random stabilizer states. These depend on the
probability distribution of entanglement, which gives the probability of finding a certain entanglement
in a stabilizer state that has been picked at random. Therefore the objective here is to find and study the
entanglement probability distribution of randomised stabilizer states.

The paper is organized as follows. In section 2 we provide some
basic results concerning random states, define
stabilizer states, the basic technical tools and results that will be used
subsequently. In section 3 we state and prove our main result.
We study the probability distribution of the entanglement across any given bipartite split of the system.
We provide a compact, explicit and exact formula for this probability
distribution and present its proof. In section 4 we use this
result to study the average entanglement of a set of spins versus
the rest and demonstrate that this distribution
implies a concentration of measure, i.e. for large numbers of
particles the probability that the entanglement of a specific state
will deviate from the mean value decreases exponentially with that deviation.

\section{Basic techniques and definitions}
\noindent
In the following we present some basic results concerning general
random quantum states as well as basic tools for the description
of entanglement in stabilizer states.

\subsection{Entanglement of randomised quantum systems}
\noindent
Consider a system of $N$ spin-$1/2$ particles and pick random
quantum states from the unitarily invariant distribution. This may
for example concern, in an idealized setting, a gas of two-level
atoms. As the atoms collide and interact at random their energy
levels become entangled. Asymptotically the distribution on pure states becomes uniform, such that any
pure state is equally likely. For such a distribution of states we may
ask: What is the average entropy of entanglement $\mathbbm{E} S_A(N_A,N_B)$ of a set of $N_A$ spins?
Some of the first studies of this question were \cite{lubkin,lloyd} and the explicit solution('Page's conjecture')
was conjectured in \cite{page} and proven in \cite{foong}. The explicit solution is given by
\begin{equation}
\label{eq:page}
\mathbbm{E} S_A(N_A,N_B)=\frac{1}{ln2}\left(\sum_{k=2^{N_B}+1}^{2^{N_A+N_B}}\frac{1}{k}-\frac{2^{N_A}-1}{2^{N_B+1}}\right)
\end{equation}
with the convention that $N_A \leq N_B$ and where $N_A+N_B=N$, the total number of particles.

This can be used to show that the average entanglement is very
nearly  maximal, meaning close to $N_A$, for large quantum systems, i.e.
$N\gg 1$. Hence one concludes that a randomly chosen state will
be nearly maximally entangled with a large probability. Indeed, it
was recently shown that the probability that a
randomly chosen state will have an entanglement $E$ that deviates by
more than $\delta$ from the mean value $\mathbbm{E} S_A(N_A,N_B)$
decreases exponentially with $\delta^2$. This phenomenon,
alongside many other properties, which is known as concentration
of measure of entanglement, was proven and extensively studied in
\cite{hayden}.

\subsection{Entanglement of stabilizer states}
\noindent
As it is our aim to study the typical properties of stabilizer
states we use this subsection to present a number of basic tools
and observations that are useful in this context.

Stabilizer states are a discrete subset of general quantum states,
which can be described by a number of parameters scaling
polynomially with the number of qubits in the state
\cite{gottesman1,gottesman2,NC}.

A {\em stabilizer operator} on $N$ qubits is a tensor product of
operators taken from the set of Pauli operators
\begin{equation}
X:=\left(\begin{array}{cc}0&1\\1&0\end{array}\right),\,\,
Y:=\left(\begin{array}{cc}0&-i\\i&0\end{array}\right),\,\,
Z:=\left(\begin{array}{cc}1&0\\0&-1\end{array}\right),
\end{equation}
and the identity $I$. An example for $N=3$ would be the operator
$g=X\otimes I \otimes Z$. A set $G=\{g_1,\ldots,g_K\}$ of $K$
mutually commuting stabiliser operators that are independent,
i.e.\ $\prod_{i=1}^{K} g_i^{s_i}=I$ exactly if all $s_i$ are
even, is called a {\em generator set}. For $K=N$ a generator
set $G$ uniquely determines a single state $|\psi\rangle$ that
satisfies $g_k|\psi\rangle=|\psi\rangle$ for all $k=1,\ldots,N$.
Such a generating set generates the stabilizer group.
Each unique such group in turn defines a unique {\em stabilizer state}.

A first observation that will be useful for the following
considerations is the fact that the bipartite entanglement of a
stabilizer state, i.e. the entanglement across any bipartite split, takes only integer
values \cite{Audenaert P 05,fattal}.

In the proof of our main theorem in section III below we will furthermore use results from
\cite{fattal}. In particular we use their result 1, that a
stabilizer state of $E$ ebits can be generated from:
\begin{itemize}
    \item {'local generators' of type $g_A\otimes I_B$ and
    $I_A\otimes g_B$ where g refers to a member of the Pauli group.}
    \item{'non-local generators' of type $g_A\otimes g_B$.
    These generators come in pairs where the entries corresponding
    to Alice (and Bob respectively) anti-commute. }
\end{itemize}
The subgroups generated by the local elements are labelled ${\cal
S}_A$ and ${\cal S}_B$ respectively, and the subgroup generated by
the non-local elements ${\cal S}_{AB}$. The entanglement $E$ of a
state is given by the number of pairs in the minimal generator set
of all non-local pairs, i.e.
\begin{equation}
E = |{\cal S}_{AB}|/2.
\label{eq:sab}
\end{equation}
For example the GHZ state $\ket{000}+\ket{111}$ with respect to
the division $N_A=2$ (first two particles) and $N_B=1$ is defined
by the generator set $\langle XXX, IZZ, ZZI \rangle$. The local
generators on Alice's and Bob's part respectively are given by
$\langle ZZI \rangle$ and $\langle I \rangle$. The non-local part
is $\langle XXX, IZZ \rangle$ which consists of only one pair so
that the entanglement is $E=1$. We will also use Eq. 6 from
\cite{fattal} giving entanglement $E$ as
\begin{equation}
\label{eq:E}
E=N_A-|S_A|=N_B-|S_B|
\end{equation}
where the $|.|$ refer to the size of the minimal generator set of
the subgroup in question.

Finally we use the fact that in order for the stabilizer state to be non-trivial it
is necessary and sufficient that the elements of the stabilizer group (a) commute,
and (b) are not equal to {\em-I} \cite{NC}.

\section{Main Result- Entanglement Probability Distribution}
\noindent
Let us consider the setting shown in fig. 1 where a system of $N$
spin-$\frac{1}{2}$ particles is split into two arbitrarily
sized subsets consisting of $N_A$ and $N_B$ particles such
that $N=N_A+N_B$. Consider now a randomly chosen pure
stabilizer state on the $N$ particles which is chosen according to
the uniform measure.
\begin{figure}[th]
\centerline{
\includegraphics[width=7.5cm]{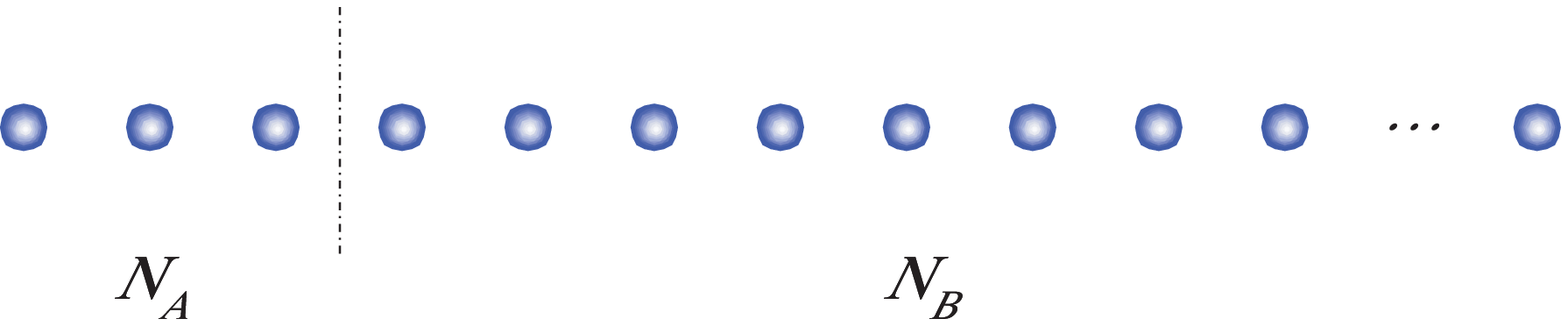}
}
\label{fig:notation}
\vspace{.0cm} \fcaption{The spins are grouped into two sets $A$ and
$B$. Set $A$ contains $N_A$ spins, set $B$ contains $N_B$ and the
total number of spins is $N=N_A+N_B$. }
\end{figure}
Generally the spins in set $A$ will exhibit entanglement with the
spins in set $B$ and this entanglement will take discrete values
$E\in\{0,1,2,\ldots,min(N_A,N_B)\}$. The main result of this paper is an
exact and compact expression for the probability distribution of
the entanglement values $E$ under randomised stabilizer states.

{\bf Theorem I --}{\em Entanglement probability distribution:}
In a system of $N$ spins where $N_A$($N_B$) is Alice's (Bob's) number of qubits the probability of finding that entanglement between $A$ and $B$ equals $E$ in a randomly chosen pure stabilizer state is given by
\begin{eqnarray*}
P(E)= \frac{\prod_{i=1}^{N_A}(2^i+1)}{\prod_{k=N-N_A+1}^{N}(2^k+1)}
\prod_{j=1}^{E}
\frac{ \left(2^{N-N_A+1-j}-1\right)\left( 2^{N_A+j}- 2^{2j-1}\right)}
{2^{2j}-1 }
\label{eq:probdist}
\end{eqnarray*}
where $E$ is an integer and $0 \leq E \leq min(N_A,N_B)$.

{\em Proof:}
The strategy of the proof is to firstly define the
probability distribution on stabilizer states. Then we count the number of
states for which the entanglement between sets $A$
and $B$ is given by $E$ and label this number by $n_E$.
Finally the probability weight is multiplied by
$n_E$ to obtain the probability distribution of entanglement.

For a set of $N$ spins we denote the {\em total number of possible
stabilizer states} by $n_{tot}(N)$. Randomising these states is
then achieved by the random application  of unitary maps that take
any stabilizer states to another stabilizer state. The probability
distribution that is invariant under randomisation is the uniform
distribution. Therefore the probability of picking a stabilizer
state $j$ at random is given by $p_j=\frac{1}{n_{tot}(N)}$.

For given values of $N$, $N_A$ and $N_B$, let $E$ denote the
amount of entanglement between set $A$ and $B$, and let
$n_E^{(N_A)}(N)$ denote the number of stabilizer
 states on $N$ particles realizing this value of entanglement.
It will be our task to determine $n_E^{(N_A)}(N)$ exactly.
It is clear that for any $N_A$ it must be the case that $n_{tot}(N)$ satisfies:
\begin{equation}
    n_{tot}(N) = \sum_{E=0}^{\lfloor N/2\rfloor} n_E^{(N_A)}(N) \, 
\label{eq:nxsum}
\end{equation}
where $\lfloor x\rfloor$ denotes the largest integer smaller than $x$.

The proof of our theorem will now be presented in the form
of several useful lemmas.\\

{\bf Lemma I --}{\em Total number of states:} The total number of distinct stabilizer states
for $N$ particles is given by
\begin{equation}
    n_{tot}(N) = 2^N \prod_{k=1}^{N} (2^k+1)\,.
    \label{eq:ntot}
\end{equation}
\\
{\em Proof:} This is proven in \cite{gottesman3} and an alternative proof
employing different techniques is found in \cite{David}\square\,
\\

{\bf Lemma II --}{\em Number of separable states:}
The number of pure stabilizer states for which the spins of sets $A$ and $B$ are {\em not} entangled is given by
\begin{equation}
    n_{0}^{(N_A)}(N) = n_{tot}(N_A)n_{tot}(N_B).\\
    \label{eq:lemmaII}
\end{equation}
{\em Proof:} A pure state does not exhibit entanglement between the spins
of sets $A$ and $B$ exactly if it is a tensor product between the two sets.
Therefore the total number of unentangled pure stabilizer states is the product of the total number of states on each part\square\,\\

{\bf Lemma III --}{\em Invariant ratio of probability
distribution:} For $0\leq E \leq min(N_A,N_B)$ and $N=N_A+N_B$ we have
\begin{equation}
    \frac{n_{E}^{(N_A+1)}(N+1)}{n_{E}^{(N_A)}(N)} =
    \frac{n_{E}^{(N_A+1)}(N_A+1+E)}{n_{E}^{(N_A)}(N_A+E)}\,.
\label{eq:constratA}
\end{equation}

{\em Proof:} This is true exactly if there is, for
arbitrary $N_A$ and $0\leq E \leq min(N_A,N_B)$, a constant ratio
\begin{equation}
\frac{n_E^{(N_A+1)}(N+1)}{n_E^{(N_A)}(N)}=f(E,N_A)
\label{eq:constrat}
\end{equation}
where $f(E,N_A)$ is some function that only depends on $E$ and $N_A$ but
{\em not} on $N_B$. To see this we now employ the methods described in
section II. Note firstly that the numbers $n_E$ are defined by counting how many
 stabilizer states there are corresponding to
a given entanglement. This is given by the number of distinct
stabilizer groups with that entanglement. For this number to be non-zero
we only consider counting it when $0\leq E \leq min(N_A,N_B)$.
The stabilizer groups have to be Abelian and contain no negative
identity, as mentioned in the introduction. Consider now the
number of ways of picking stabilizer group elements, under these
restrictions, on Bob's side. We can call this number $\beta$ and
the corresponding number on Alice's side $\alpha$. The problem is
simplified by counting the number of distinct generator sets of
the group. Several generator sets will generate a given group, so
the number of generator sets is larger than the number of groups
by some overcount factor. We now show that $\beta$, as well as the
overcount factor, are invariant during the transition $\tau:N_A
\mapsto   N_A+1$ under constant entanglement $E$.

Firstly note that the contribution of Bob's side to $\beta$, and
to the overcount factor, is independent of $|S_A|$. This is
because there are only identities in the corresponding terms on
Bob's side. It does depend on $|S_B|$ and $|S_{AB}|$ though.
Eq. \ref{eq:E} implies that $|S_A|$, but not $|S_B|$ changes
under the transition $\tau:N_A \mapsto   N_A+1$.  The condition of
constant $E$ is equivalent to constant $|S_{AB}|$ by Eq.
\ref{eq:sab}.

So  $\tau:\beta\mapsto \beta$ and $\tau:\alpha\mapsto \alpha'$. Therefore Bob's contribution
to the ratio (\ref{eq:constrat}) cancels; the ratio is independent of $N_B$. \square\,\\

Now we note that the following closed form expressions
\begin{equation}
    n_E^{(N_A)}(N) =
    \left(2^{N+1-N_A-E}-1\right)\frac{2^{N_A+E}-2^{2E-1}}
    {2^{2E}-1}n_{E-1}^{(N_A)}(N)
\label{eq:nj}
\end{equation}
satisfy equations (\ref{eq:nxsum}), (\ref{eq:ntot}), (\ref{eq:lemmaII}) and (\ref{eq:constratA}),  for $0\leq E \leq min(N_A,N_B)$, i.e. for all possible settings.
What remains to be shown is that this choice is indeed unique, i.e. that we can construct $n_{E}^{(N_A)}(N)$ recursively
and uniquely from equations (\ref{eq:nxsum}), (\ref{eq:ntot}), (\ref{eq:lemmaII}) and (\ref{eq:constratA}).
We proceed in the following steps.
\begin{itemize}
\item[$\bullet$] Note that we know
$n_{E}^{(N_A)}(N)$ for $N_A=1$ and all valid choices
of $N$ for $E=0$ by virtue of Lemma II and
then for $E=1$ by virtue of eq. (\ref{eq:nxsum}).

\item[$\bullet$] Assume that we have obtained
$n_{E}^{(N_A)}(N)$ for $N_A=1,2,\ldots,r_0$ and all
valid choices of $E\in\{0,\ldots,N_A\}$ and
$N\in\{N_A+E,\ldots\}$. Now we will demonstrate
that this uniquely defines $n_{E}^{(r_0+1)}(N)$
for all $E\in\{0,1,\ldots,r_0+1\}$
and all $N\in\{r_0+1+E,\ldots\}$.

\begin{itemize}
\item[$\bullet$] To this end realize first that $n_{E=0}^{(r_0+1)}(N)$
is again known for all $N$ by virtue of Lemmas I and II. Now
assume that for fixed $r_0+1$ we have found $n_{x}^{(r_0+1)}(N)$
for $x\in\{0,1,\ldots,E\}$ and all $N$.

\item[$\bullet$] We find $n_{E+1}^{(r_0+1)}(N)$ for all
$N$ in the following way. First we use eq.(\ref{eq:nxsum}) to
obtain $n_{E+1}^{(r_0+1)}(E+r_0+2) = n_{tot}(E+r_0+2) -
\sum_{j=0}^{E} n_{j}^{(r_0+1)}(E+r_0+2)$. Now we employ the
recursion relation in Lemma III to obtain
$n_{E+1}^{(r_0+1)}(N+1) = \frac{n_{E+1}^{(r_0+1)}(E+r_0+2)}
{n_{E+1}^{(r_0)}(E+r_0+1)}n_{E+1}^{(r_0)}(N)$ where each
term on the right hand side is already known by assumption.
\end{itemize}

This completes the construction for arbitrary $N$ which
in turn completes the construction for arbitrary $E$.
Therefore, all $n_{E}^{(N_A)}(N)$ are uniquely determined.
It is now cumbersome but straightforward to check that
eq.(\ref{eq:nj}) satisfies all the recursion relations
which in turn implies that it is the unique solution
presenting the correct values for $n_{E}^{(N_A)}(N)$.

\end{itemize}

This finishes the proof for Theorem I \square\,

\section{Implications}
\noindent
This section will present a discussion of the implications
of Theorem I.

\subsection{Product-free form}
\noindent
Theorem I can be more conveniently evaluated in the
format free from products
\\
{\bf Corollary I --}{\em Product-free form:} The
probability of $E$ entanglement is
\begin{equation}
    P(E)=2^{ \frac{(N_A-N_B)^2}{4}-(N/2-E)^2 +\Sigma _1 +\Sigma _2 }
\label{eq:productfree}
\end{equation}

\noindent
where $N_A$, $N_B$ and $N$ are as defined earlier, and\\
\begin{equation}
 \Sigma _1= \sum_{j_1=1}^{N_A}log_2(1+2^{-j_1}) +
 \sum_{j_2=1}^{N-N_A}log_2(1+2^{-j_2}) -
  \sum_{j_3=1}^{N}log_2(1+2^{-j_3})
\end{equation}
and finally \\
\begin{equation}
 \Sigma _2= \sum_{k=1}^{E} log_2(1-2^{k-1-N_A}) + log_2(1-2^{-N+N_A-1+k}) - log_2(1-2^{-2k})\,.
\end{equation}
{\em Proof:} This is a reexpression of Theorem I\,\square\\
We will see $\Sigma _1$ and $\Sigma _2$ converge to numbers of
magnitude $\ll$ $N$ when $N$ is large so they can be seen as minor modulations to the leading Gaussian-type behaviour. This statement will be made more
rigorous later on in this section.

\subsection{Comments on entanglement probability distribution}
\noindent
It is instructive to consider qualitatively what kind of
distributions Theorem I describes, before making precise
mathematical arguments. Firstly consider the simplest non-trivial case, that
of $N$=2, $N_A=N_B=1$. Here one obtains that the number of states
with one ebit is $n_1=24$, whereas $n_0=36$ with $n_{tot}=60$.
Hence the expected entanglement is $0.4$. If we now increase
Bob's size to $N_B=2$, the formulae show that $n_0=360$, $n_1=720$
with  $n_{tot}=1080$, giving an average entanglement of
$\frac{720}{1080}=\frac{2}{3}$. When considering all possible
$N_A$ under increasing $N$, one observes that the maximal
and near maximal values of entanglement become the main
contributors to the number of possible states. This can be
seen from Eq. (\ref{eq:productfree}) in which the leading
terms inside the sum correspond to a Gaussian centred on $N/2$.
Note however that although the corrections to this Gaussian,
such as the $\Sigma_2$ term are small, they contain an $E$-dependence
that adds some subtleties to this behaviour that will require a
more rigorous analysis.
This can be understood by noting from Eq. (\ref{eq:nj}), that
\begin{equation}
\frac{n_{N_A-1}}{n_{N_A}}=\frac{1}{2^{N-2N_A}-\frac{1}{2}}-\frac{1}{2^N-2^{-2N_A-1}}
\label{eq:namax}
\end{equation}
where the second term quickly disappears with increasing $N$. If $N_A$ is constant, the ratio tends to 0 exponentially fast.
However if $N_A=\frac{N}{2}$ then
$n_{N_{A-1}}/n_{N_A}\rightarrow 2$ for large $N$ which implies in fact
that the maximum of the probability distribution is shifted to $N/2-1$.
In that case one sees from Eq.
(\ref{eq:nj}) that $n_{N_{A-2}}/n_{N_{A-1}}\rightarrow \frac{2}{9}$, after which the ratios are very small. Hence for this case $N_A-1$ is the most likely entanglement of a state picked at random rather than $N_A$.

We can now use the entanglement probability distribution to derive the average entanglement of randomised stabilizer states.

\subsection{Average entanglement of randomised stabilizer states}
\noindent
{\bf Corollary II --}{\em Average entanglement:}
The average entanglement, $\mathbbm{E}S_A$, in stabilizer states sampled at random is given by
\begin{eqnarray*}
        &\mathbbm{E}S_A=\frac{\prod_{i=1}^{N_A}(2^i+1)}
        {\prod_{k=N-N_A+1}^{N}(2^k+1)}
        \sum_{E=1}^{N_A} E \prod_{j=1}^{E}
        \frac{ \left(2^{N-N_A+1-j}-1\right)\left( 2^{N_A+j}-
         2^{2j-1}\right)}{2^{2j}-1 }
\end{eqnarray*}
where $N_A$ is the number of qubits belonging to Alice and $N$ is the total number of qubits. We follow, without loss of generality,
the convention that $N_A \leq N-N_A=N_B$. The total state is bipartite and pure and
the average is taken over a flat distribution on stabilizers.

{\em Proof:} We take as a starting point
\begin{equation}
    \mathbbm{E}S_A = \sum_{E=1}^{N_A}P(E)E
\label{eq:startpt}
\end{equation}
which is then evaluated using Theorem I\,\square

It is also worth mentioning that the case $N_A=1$ yields the simple form
\begin{equation}
\mathbbm{E}S_A=\frac{n_{tot}-n_0}{n_{tot}}=1-\frac{3}{2^N+1}.
\label{eq:na1}
\end{equation}

\subsection{Concentration of measure}
\noindent
If some property becomes very likely in certain
circumstances, we say the measure is becoming
concentrated around states with this property.
It is known\cite{hayden} that the measure on
general pure states is concentrated around
states with near maximal entanglement. Here we
show that a similar concentration takes place on
stabilizer states, with some subtleties. We know
that the mean is close to maximal which we can use
as an upper bound. Below we give an exact expression for the probability
of picking a state with an entanglement less than the average by some number.
We then give an inequality for the same quantity.

{\bf Corollary III --}{\em Concentration of measure
around mean in stabilizer states:}
The probability of picking a stabilizer state such that it is
at least $\varepsilon$ less than the average $\mathbbm{E}S_A$
is given by
\begin{equation}
        P(S_A < \mathbbm{E}S_A-\varepsilon )=\sum_ {E=1}^{\lfloor \mathbbm{E}S_A-\varepsilon \rfloor}P(E)
\label{eq:concentration}
\end{equation}
where P(E) is given by Theorem I and $\mathbbm{E}S_A$ by Corollary II.
$\varepsilon$ is a constant which can take any value in the range of 1 to $N_A-1$.

The concentration of measure can also be given in the shape of an inequality by exploiting the fact that the leading behaviour of the probability distribution of entanglement is Gaussian:
\begin{equation}
\sum_{E=1}^{\lfloor \mathbbm{E}S_A-\varepsilon \rfloor}P(E)  \leq
\gamma (N,N_A)\int _{1}^{1+\lfloor \mathbbm{E}S_A-\varepsilon \rfloor} dEe^{-ln2(N/2-E)^2}
\label{eq:concentrationbound}
\end{equation}
where
\begin{equation}
\gamma (N,N_A)= 2^{\frac{\left( N_A - N_B\right)^2}{4}+s_1+s_2}\,;
\end{equation}
$s_1$ is defined in Eq. (\ref{eq:s1}). $s_2$ is given by
\begin{equation}
s_2=\frac{h^{N_A}+h^{N_B}}{1-h}-log_2(15/16)+h
\end{equation}
where for typographical reasons $-log_2(3/4)=h$.

{\em Proof: } The terms inside the integral are taken from the product free form of the probability distribution. We then factor out
$\gamma$. This is possible by using $E$-independent bounds on $\Sigma_1$ and $\Sigma_2$. The latter is an $E$-independent version
of Eq. \ref{eq:s2} below. The integral can be taken as a bound to the sum, as the remaining Gaussian is centred on $N/2$, and without loss of generality $N_A \leq N/2$, meaning there is a monotonic rise within the bounds
of the integral\,\square

The Gaussian integral can be evaluated using the error function.

\subsection{Comparison with general random states}
\noindent
The typical entanglement of stabilizer states is very similar to that of general states. By this statement we refer to two properties of the entanglement in random
large general states picked from the unitarily invariant distribution:
the average entanglement is near maximal, $N_A$, and the probability distribution is exponentially concentrated around
this average.

The concentration of measure section above shows that the latter is true also for random large stabilizer states. The typical entanglement
of stabilizer states is nearly maximal, as is the case for general states. The averages are slightly different though, and the concentration of measure
around the average appears to be a bit less abrupt in the stabilizer state case.

Figure 2 shows how the averages of randomised general states and stabilizer states compare for the full range
of possible Alice and Bob divisions where the total system is ten qubits.

\begin{figure}[th]
\centerline{
\includegraphics[width=12.5cm]{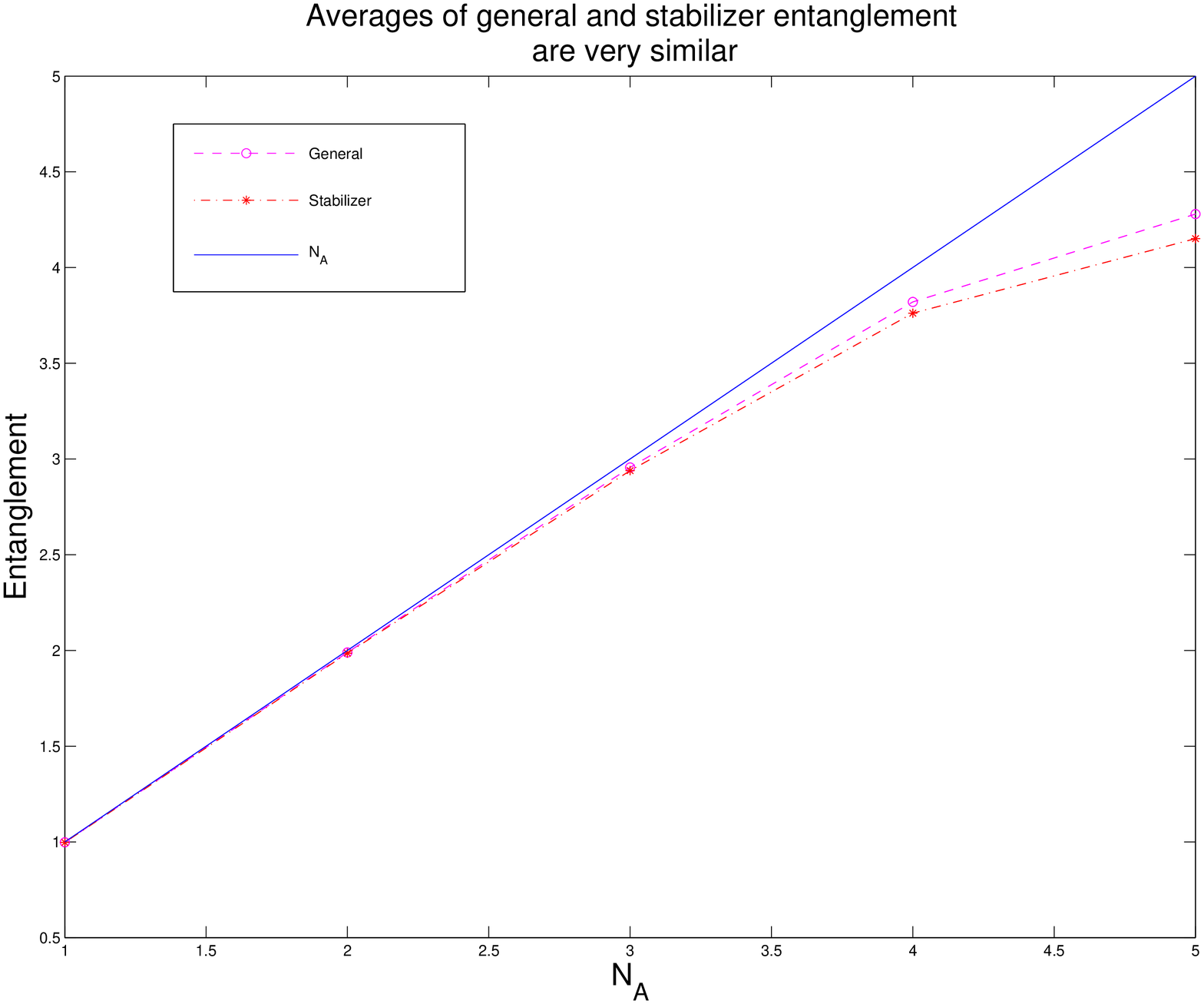}
}
\label{fig:fixedN}
\vspace{.0cm}
\fcaption{For a system of ten qubits with $N_A$ belonging to Alice the average entanglement of both general\cite{page} and stabilizer states(Corollary II) are shown.}
\end{figure}

In order to prove this similarity for arbitrary $N_A$ a good starting point is a rearrangement of Corollary II.
\begin{equation}
    \mathbbm{E}S_A =N_A(1-\frac{n_0}{n_{tot}})-
    \sum_{E=1}^{N_A-1}(N_A-E)2^{ \frac{(N_A-N_B)^2}{4}-(\frac{N}{2}-E)^2 +\Sigma _1 +\Sigma _2}
\label{eq:NAtype1}
\end{equation}
which uses the product free form of Theorem I as in Eq. (\ref{eq:productfree}).
To obtain a lower bound on this we will need the following
upper bounds on $\Sigma _1$ and $\Sigma _2$. \\
$\bullet$ {\em $\Sigma_1$ upper bound:}\\
A compact expression for $N_A=1$ was given in eq. (\ref{eq:na1})
so that we may concentrate on the case $N_A \geq 2$. Then
$\Sigma_1  \leq s_1$ where
\begin{eqnarray}
 s_1 \equiv log_2(3/2)+log_2(5/4)+\frac{13}{36} \left[1-  \left( \frac{26}{50}\right) ^{N_A-2}\right]\,.
\label{eq:s1}
\end{eqnarray}
Proof:
Firstly note
\begin{eqnarray}
\Sigma _1 &=&\sum_{j1=1}^{N_A}log_2\left(1+2^{-j1}\right)-\sum_{j3=N_B+1}^{N}log_2\left(1+2^{-j3}\right) \\
& \leq & \sum_{j1=1}^{N_A}log_2\left(1+2^{-j1}\right) \\
& = & log_2(3/2)+log_2(5/4)+\sum_{j1=3}^{N_A}log_2\left(1+2^{-j1}\right)\,.
\label{eq:sigma1upperbound2}
\end{eqnarray}
Now consider using an upper bound of the type
$exp(a+bk)$ on $log_2(1+2^{-2-k})$, where
$a$ and $b$ are constants to be determined.
Choose $a$ and $b$ such that the bound is
exact for $k=1$ and $k=2$ respectively. This
yields $a=2ln(log_2(9/8))-ln(log_2(17/16))\cong -1.10825...$
and $b=ln(log_2(17/16))-ln(log_2(9/8))\cong -0.664143...$.
That $exp(a+bk)$ is indeed an upper bound
to $log_2(1+2^{-1-k})$ for $k\ge 2$ follows
by immediately comparing the gradients
of the two functions. One
now substitutes this bound into Eq.
(\ref{eq:sigma1upperbound2}) and uses the
standard formula for sums of geometric sequences,
as well as bounding the messy-looking logarithm
ratios by rational numbers, to recover Eq.
(\ref{eq:s1})\square\,

The upper bound on $\Sigma _2$ can be derived
employing very similar ideas.\\
$\bullet$ {\em $\Sigma_2$ upper bound:}
To achieve a more compact notation let $h=-log_2(3/4)$
\begin{equation}
  \Sigma _2 \leq \frac{1-h^{-E}}{1-h}\left[ h^{N_A}+h^{N_B} \right]+
\frac{4}{3}\left( 2^{-2E}-1\right)log_2\left(15/16\right)+h \,.
\label{eq:s2}
\end{equation}

{\em Proof:}
Essentially the same method is used as in the bound for $\Sigma_1$. We use two inequalities that are valid within the range used only.
Firstly
\begin{equation}
-log_2\left(1-2^{-2j-2}\right)\leq e^{ln(-log_2(15/16))-2ln2+2ln2j}
\end{equation}
which was found using an $\exp(a+bj)$ type bound which we required
to be exact for $j=1$. The maximum allowable gradient such that
it is an upper bound for $j\geq1$ was used. Secondly
\begin{equation}
log_2\left(1-2^{-x}\right)\leq - e^{ln(-log_2(3/4))(1-x)}
\end{equation}
which was found using an $\exp(a+bx)$ type bond, where a and b are
chosen such that the bound is exact for $x=1$ and $x=2$. The
gradients of the two functions are such that it will be an
upper bound for all points in the range. These two bounds
are then inserted into the sum in the definition of $\Sigma_2$
and with the standard formula for sums of geometric sequences
one recovers Eq. (\ref{eq:s2})\,\square

Using these bounds in Eq. (\ref{eq:NAtype1}) provides a
very tight fit to the exact result.

\section{Summary and Outlook}
\noindent
In this paper we have studied the entanglement
properties of randomised stabilizer states on $N$
particles. We considered a bi-partite setting with
two sets containing $N_A$
and $N_B$ spins with $N=N_A+N_B$ and the entanglement
between them. We obtain an exact and compact probability
distribution for the possible values of entanglement
(which are integers) and use this to give exact values for the average
entanglement and the concentration of measure around
it with increasing size of the system. It was found
that, for large systems, the average entanglement is
nearly maximal and the probability of a state having a
significantly different entanglement than this average
decreases exponentially with the difference.
We also compared these results to the case of general
random quantum states and found a close similarity.

This suggests that it would be interesting to conduct
research into how far this similarity carries, in particular
in the context of multi-partite entanglement between
large and/or non-contiguous sets of particles and of
dynamical features of quantum states under the action
of random quantum gates (stabilizer or general unitary gates)
\cite{lloyd,Calsamiglia HDB 05}
and their approach to the equilibrium distribution.
These topics will be the subject of a forthcoming
publication currently in preparation.

\nonumsection{Acknowledgements}
\noindent
We are grateful to K. Audenaert, J. Eisert, D.
Gross, J. Oppenheim, T. Rudolph, G. Smith for discussions and
especially to S.Virmani as well as an unknown referee for carefully reading the manuscript. This
work is part of the QIP-IRC supported by EPSRC (GR/S82176/0) and
was supported by The Leverhulme Trust, The Royal Society and the
European Union Integrated Project QAP.\\

{\em Note added:} After completion of this research we became aware of
work by Graeme Smith and Debbie Leung, \cite{Leung}, which
considers the same issues and some additional questions but does
not provide exact expressions for the probability distribution for
the entanglement of a set of spins versus the rest as we do here.

\nonumsection{References}


\noindent
\begin{thebibliography}{99}
\bibitem{Plenio V 98} M.B. Plenio and V. Vedral (1998), {\em Teleportation, Entanglement and Thermodynamics in the Quantum World}, Contemp. Phys.
{\bf 39}, 431.
%
\bibitem{Eisert P 03} J. Eisert and M.B. Plenio (2003), {\em Introduction to basics of Entanglement Theory in Continuous Variable Systems }, Int. J. Quant. Inf. {\bf 1}, 479.
%
\bibitem{Plenio V 05} M.B. Plenio and S. Virmani (2005), {\em An Introduction to Entanglement Measures }, E-print arxiv
quant-ph/0504163.
%
\bibitem{Eisert G 05} J. Eisert and D. Gross (2005), {\em Multi-Particle Entanglement }, E-print arxiv
quant-ph/0505149.
%
\bibitem{Horodecki1}M. Horodecki (2001), {\em Entanglement measures}, Quantum Information and Quantum Computation, {\bf 1,1}  (pp3-26)
%
\bibitem{Wootters}W. Wootters (2001), {\em Entanglement of formation and concurrence}, Quantum Information and Quantum Computation, {\bf 1,1}  (pp27-44)
%
\bibitem{Horodecki2}P. Horodecki, R. Horodecki (2001), {\em Distillation and bound entanglement},  Quantum Information and Quantum Computation, {\bf 1,1}  (pp45-75)
%
\bibitem{gottesman1} D. Gottesman, {\em Stabilizer Codes and Quantum Error Correction},
Caltech PhD thesis.
%
\bibitem{gottesman2}D. Gottesman (1998), {\em The Heisenberg Representation of Quantum
Computers}, E-print arxiv quant-ph/9807006.
%
\bibitem{Audenaert P 05} K.M.R. Audenaert and M.B. Plenio (2005), {\em Entanglement on mixed stabiliser states: 
Normal Forms and Reduction Procedures}, New J.
Phys. {\bf 7}, 170. Note that the Matlab codes in this work
can be downloaded from www.imperial.ac.uk/quantuminformation.
%
\bibitem{gottesman3} D.Gottesman and S.Aaronson (2004),
{\em Improved Simulation of Stabilizer Circuits}, Phys. Rev.A, 70:052328.
%
\bibitem{fattal} D. Fattal, T. S. Cubitt, Y. Yamamoto, S. Bravyi and I.L. Chuang (2004), {\em Entanglement in the stabilizer formalism}, E-print arxiv
quant-ph/0406168.
%
\bibitem{Hein EB 04} M. Hein, J. Eisert and H.J. Briegel (2004), {\em Multiparty Entanglement in Graph States},
Phys. Rev. A {\bf 69}, 062311. 
%
\bibitem{lubkin} E.Lubkin (1978), {\em Entropy of an n-system from its correlation with a k-reservoir},
J. Math. Phys {\bf 19}(5), pp.1028-1031 .
%
\bibitem{page} D.N.Page (1993), {\em Average entropy of a subsystem}.  Phys. Rev. Lett. {\bf 71} No.9, (1993).
%
\bibitem{foong} S.K Foong and S.Kanno (1994), {\em Proof of Page's conjecture on
the average entropy of a subsystem}.  Phys. Rev. Lett. {\bf 72}
No. 8, pp.1148-1151.
%
\bibitem{lloyd} S.Lloyd and H.Pagels (1988), {\em Complexity as Thermodynamic Depth.}
 Ann. of Phys. {\bf 188}, 186-213.
%
\bibitem{hayden} P.Hayden, D.W.Leung and A.Winter (2004),
{\em Aspects of Generic Entanglement}, Comm. Math. Phys. March 2006.
%
\bibitem{Leung} G.Smith and D. Leung (2005), {\em Typical Entanglement of Stabilizer
States}, E-print arxiv quant-ph/0510232.
%
\bibitem{David} D. Gross (2006), {\em Hudson's Theorem for finite-dimensional quantum
systems}, E-print arxiv quant-ph/0602001.
%
\bibitem{Calsamiglia HDB 05} J. Calsamiglia, L. Hartmann, W. D\"ur,
H.-J. Briegel (2005), {\em Spin Gases: Quantum Entanglement Driven By
Classical Kinematics}, Phys. Rev. Lett. {\bf 95}, 180503. 
%
\bibitem{NC} M.A. Nielsen and I.L. Chuang, {\em Quantum computation and
Quantum Information},  Cambridge Univ. Press, Cambridge, UK.
%
\end{thebibliography}
\end{document}